\documentclass[runningheads]{llncs}
\usepackage{graphicx}
\usepackage{hyperref}
\begin{document}
\title{RERS-Fuzz : Combining Greybox Fuzzing with Interval Analysis for error reachability in reactive softwares}
\titlerunning{RERS-Fuzz}
%
\author{Animesh Basak Chowdhury}
\authorrunning{Animesh et al.}
%
\institute{Indian Statistical Institute, Kolkata, India \\
\email{animeshbchowdhury@gmail.com}\\}
\maketitle              
\begin{abstract}
Fuzz Testing is a well-studied area in the field of Software Maintenance and Evolution. In recent years, coverage-based Greybox fuzz testing has gained immense attention by discovering critical security level and show-stopper bugs in industrial grade software. Greybox fuzz-testing uses coverage maximization as objective function and achieve the same by employing feedback-driven evolutionary algorithms. In our work, we have utilized the power of Greybox fuzz testing, combined with interval analysis for solving reachability problem in \texttt{sequential} and \texttt{industrial} RERS \texttt(Rigorous Examination of Reactive Software) 2019 benchmarks. 

\end{abstract}
\section{Introduction}

RERS-Fuzz is an automated test-generation tool targeted for coverage maximization in reactive softwares. The techniques employed are heavily based on key concepts of Greybox fuzz-testing~\cite{ref_greybox}. The underlying idea is to employ evolutionary algorithms for generating interesting test-inputs that help in exploration of newer code segments. In Greybox fuzz-testing, a set of random test-inputs are considered as initial population. The fitness function is the code coverage measure observed by an individual test-input. For every test-input, fitness value is calculated and best-fit test-inputs are retained for further fuzzing. The retained test-inputs are subjected to various mutation and crossover operations for generation of newer test-inputs. The cycle repeats until certain user-defined goals are met or statistical measures on coverage metrics are achieved.

We have mainly used the following tools and techniques and modified them inline with our requirements to solve reachability problems on \texttt{RERS} 2019 benchmarks :-
\begin{itemize}
\item American Fuzzy Lop (AFL v2.52b)~\cite{ref_afl}.
\item Instrim~\cite{ref_instrim}.
\item VeriFuzz~\cite{ref_verifuzz}
\item LLVM Interval Analysis~\cite{ref_intervalAnalysis}.
\end{itemize}

\noindent In subsequent sections, we briefly describe about our approach, highlighting core techniques, tool architecture and strengths and weaknesses associated with them.

\section{Approach}

State-of-art greybox fuzzers, like \texttt{afl-fuzz} uses a stronger notion of structural code coverage of program's control-flow-graph (\texttt{CFG}), called \textit{branch-pair} coverage as fitness function.  The tool maintains a shared 64kB memory where each byte entry represents a logarithmic visit count of a typical branch-pair by all test-inputs generated so far. The amount of shared memory is designed in such a way that entire shared memory can be stored in cache-memory and hence execution speed of fuzz-testing tool remains unhampered. However, as per our observation, the number of conditional statements in \texttt{RERS} benchmarks are typically very high. Hence, instrumenting each conditional statement of program would lead to heavy collision~\cite{ref_afl_whitepaper}. Secondly, the ranges of input values driving the reactive softwares is comparatively small. Hence, application of heavy-weight and sophisticated mutation and crossover operations would amount to generation of useless test-inputs rejected by the reactive system. Finally, in reactive softwares, code-coverage does not necessarily ensure that newer states are also covered.

In our tool, we have tried of identify the problems where plain usage of greybox fuzz testing tool would fail. We have added our own techniques on top of core-fuzzing engine to suitably tune the application of evolutionary fuzz testing on \texttt{RERS} benchmarks. The following are the key techniques of our approach.
\\ \\
\textbf{Minimal Node Instrumentation}: As discussed earlier, the software-under-test (\texttt{SUT}) is subjected to node instrumentation for measuring code coverage during a test-run. However, in \texttt{RERS} benchmarks the number of conditional statements are significantly high and complex. In order to reduce cache collision and execution overhead arising from instrumentation, number of instrumentation points have to be reduced. In our tool, we have used lightweight instrumentation~\cite{ref_instrim} which is efficient and instruments minimal conditional statements of \texttt{SUT}, without loosing any coverage information. The technique have posed the minimal instrumentation as \textit{path differentiation problem} and identify minimal number of nodes in \texttt{CFG} for which any two branch-pairs can be differentiated.
\\ \\
\textbf{Interval Analysis assisted Mutation}: Reactive softwares belong to class of program where input space is very restricted~\cite{ref_verifuzz}. It is possible to tune mutation and crossover operation within the ranges of input taken by ~\texttt{SUT}. We have taken help of LLVM's interval analysis~\cite{ref_intervalAnalysis} to identify ranges of input consumed by \texttt{RERS} to go into deeper program segments. The values are then passed on to our parametric fuzz engine, which in turn does crossover and mutation operation over those restricted ranges.
\\ \\
\textbf{State Instrumentation}: The minimal node instrumentation takes care of efficient branch-pair coverage in program's \texttt{CFG}. But, it may happen that a test-input have exhibited no new branch-pair coverage in its run, but it have explored a new \textbf{\textit{state}}. We define \textbf{\textit{state}} as, unique values assignments to all the global variables in \texttt{SUT}. In addition to existing 64k shared memory for branch-pair instrumentation, we have additionally deployed 64k shared memory, which maintains a binary valued entry whether a \textbf{\textit{state}} has been explored or not. Accordingly, our fitness function is also updated with state coverage information. Any test-input with no new branch-pair coverage but new state space coverage are now retained for further fuzzing.

\section{Tool Implementation}

\begin{figure}
\includegraphics[scale=0.8]{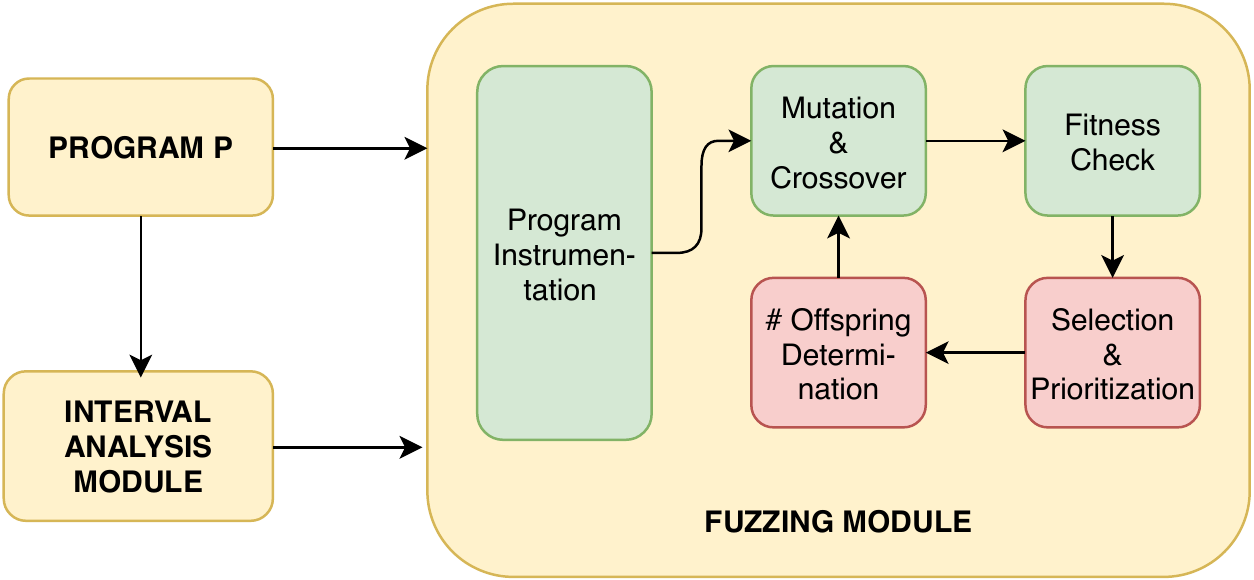}
\caption{RERS-Fuzz : Overview}
\label{fig:rersFuzzOverview}
\end{figure}

The overall flow of RERS-Fuzz has been shown in figure~\ref{fig:rersFuzzOverview}. The fuzzing module has been developed on top 
greybox fuzzer \texttt{afl-fuzz}~\cite{ref_afl}. The green blocks in the fuzzing module denotes that techniques have been modified and tuned with \texttt{RERS} programs. Red blocks denote the algorithm is unchanged and in line with afl-fuzz core algorithm. We have additionally used Instrim\footnote{https://github.com/csienslab/instrim} package for minimal node instrumentation on top of LLVM-7 framework. For Interval Analysis module, we used the LLVM package developed for interval analysis\footnote{http://llvm.org/doxygen/Interval\_8h\_source.html}. The tool has been developed using C, C++ and python.

\section{Strengths and Weakness}

RERS-Fuzz participated in Reachability Track of RERS 2019 competition. The tool could find out counterexamples for 217 academic reachability benchmarks. In industrial reachability track, it has emerged as overall winner, scoring 2038 points.

The core strength of RERS-Fuzz is its capability to scale up on large industrial benchmarks and find sufficient number of error reachable locations. We have run all sequential and industrial reachability benchmarks for a timeline of 8 hours on Intel i7 3720 Octa-Core Processor (2.6GHz) and presented our results. The entry for the specification with \texttt{error\_reachable} denotes that error location is reachable. Entries with \texttt{UNKNOWN} denote we are unable to conclude whether the error location is reachable or not.

Weaknesses involve the inability of the tool to generate proofs for unreachability of error location. Besides that, we have observed that running certain benchmarks for a longer time period can yield more number of error reachable locations. However there are no assurances that preserving test-inputs with newer state-space or branch-pair would definitely aid to uncover error-reachable locations during evolution. With our technique, we try to maximize a stronger notion code-coverage and state-space coverage of \texttt{RERS} benchmarks. Today, our tool lacks focused search towards reachable error location. We believe that in future, such problems can be modelled as appropriate fitness function and hence fuzzing can be made directed. Also, such high quality test-vectors would definitely aid various invariant learning algorithms, which in turn would help prove correctness of such reactive software.

\end{document}